\documentclass{optica-article}

\journal{oe}

\usepackage{siunitx}
\usepackage{amsmath}
\usepackage{lineno}
\nolinenumbers

\begin{document}

\title{Gallium arsenide whispering gallery mode resonators for terahertz photonics}

\author{Mallika Irene Suresh,\authormark{1,2,*}, Harald G. L. Schwefel\authormark{1,2}, Dominik Walter Vogt\authormark{1,3}}

\address{\authormark{1}The Dodd-Walls Centre for Photonic and Quantum Technologies, New Zealand\\
\authormark{2}Department of Physics, University of Otago, Dunedin 9016, New Zealand\\
\authormark{3}Department of Physics, University of Auckland, Auckland 1010, New Zealand}

\email{\authormark{*}mallika.suresh@otago.ac.nz} 


\begin{abstract}
As the field of terahertz (THz) photonics advances, we present a monolithic gallium arsenide (GaAs) disk-shaped whispering gallery mode resonator that has potential as a component in THz nonlinear optics. GaAs is a material with significant optical nonlinearity which can be enhanced when the crystal is shaped into a microdisk resonator. A 4-mm-disk-resonator was fabricated using single-point diamond turning and was characterised to obtain a quality (Q) factor of 2.2k at $\sim$\SI{150}{GHz} and 1.4k at $\sim$\SI{300}{GHz}. We also demonstrated the blue-shifting of up to $\sim$\SI{0.3}{GHz} of the THz modes using a block of metal. This post-fabrication degree of freedom could be useful for phase-matching requirements for nonlinear optical processes, such as detection based on optical up-conversion of THz radiation. Such a compact, tunable and efficient device could be integrated into nonlinear photonic platforms for THz generation, manipulation and detection.
\end{abstract}

\section{Introduction}
For many years, the spectral region between 0.1 - 10 THz was seen as inaccessible. Instruments used for generating or detecting radiation on either the lower frequency side (electronic) or the high frequency side (optical) were insufficient at THz frequencies, resulting in this being called the "THz gap". However, there has recently been a significant focus on the development of efficient, cost-effective and compact generation and detection of THz radiation~\cite{samizadeh_nikoo_electronic_2023, guerboukha_conformal_2023, dastrup_enhancement_2022, ojo_thz_2023,carpintero_selected_2015}. This has been propelled by the applications in communication, imaging and sensing ranging from medical to security and defense~\cite{moldosanov_terahertz_2017, lu_review_2021, fu_applications_2022, huang_terahertz_2022}. A lot of the discussion and investigation into designing and exploring materials and structures for manipulating THz radiation is fueled by the idea to use platforms and fabrication schemes that are regularly used in electronics and optics, particularly in implementing silicon (Si) photonic structures for THz~\cite{nagatsuma_advances_2016, zhou_photonics-inspired_2021, xie_review_2021}. There is also notable interest in developing integrated platforms for efficient nonlinear processes combined with a small footprint for THz sources and sensors~\cite{hou_photonic_2020, yang_terahertz_2020, tal_nonlinear_2020, carnio_nonlinear_2023}.

Here, we contribute to the field by presenting the fabrication and characterisation of a THz whispering gallery mode resonator (WGMR) made out of gallium arsenide (GaAs). We present a GaAs WGMR that is designed for use in the \qtyrange{150}{380}{GHz} spectral range. This spectral region has been of interest for applications ranging from identifying signatures of aromatic chiral molecules for interstellar searches~\cite{stahl_laboratory_2020} to demonstrating wireless communication at a high data rate~\cite{ali_300_2019, alibakhshikenari_study_2020, fujishima_future_2021}. To the best of our knowledge, GaAs has not been explored in the form of a WGMR in the THz domain before. WGMRs are useful as compact resonators in applications such as tunable filters~\cite{wang_voltage-actuated_2019} and isolators~\cite{yuan_on-chip_2021}, but they are also implemented as sensitive detectors and low-threshold sources~\cite{bravo-abad_efficient_2010, sinha_tunable_2016}, as well as for sensitive material characterisation~\cite{vogt_terahertz_2022, vogt_coherent_2019} and for fundamental studies~\cite{preu_coupled_2008,preu_directional_2013}. High-Q(uality) WGMRs provide a platform of tight confinement of the resonant fields in a compact volume for an extended period of time. This increase in the interaction between the fields and the material can lower the threshold of nonlinear optical processes, as has been demonstrated at optical frequencies~\cite{strekalov_nonlinear_2016} as well as in the THz domain~\cite{chassagneux_terahertz_2007, amiri_silicon-based_2016, masini_continuous-wave_2017,strekalov_microwave_2009,strekalov_towards_2009
}. As a crystal with second-order optical nonlinearity, WGMRs fabricated out of GaAs could be used for nonlinear optical mixing processes such as difference frequency generation to implement THz sources~\cite{vodopyanov_terahertz-wave_2006, schaar_intracavity_2007}, as well as frequency doubling and up-conversion to optical frequencies for THz detection~\cite{eakkapan_new_2012, sinha_tunable_2016, cosci_THz_2017}. In addition, as GaAs is a material that is used in the semiconductor industry, one could take advantage of existing fabrication techniques for mass manufacturing of integrated chip-scale THz hybrid systems.

\section{Experimental methods}
\subsection{Design and fabrication of the GaAs WGMR}
\begin{figure}[h!]
\centering\includegraphics[width=12cm]{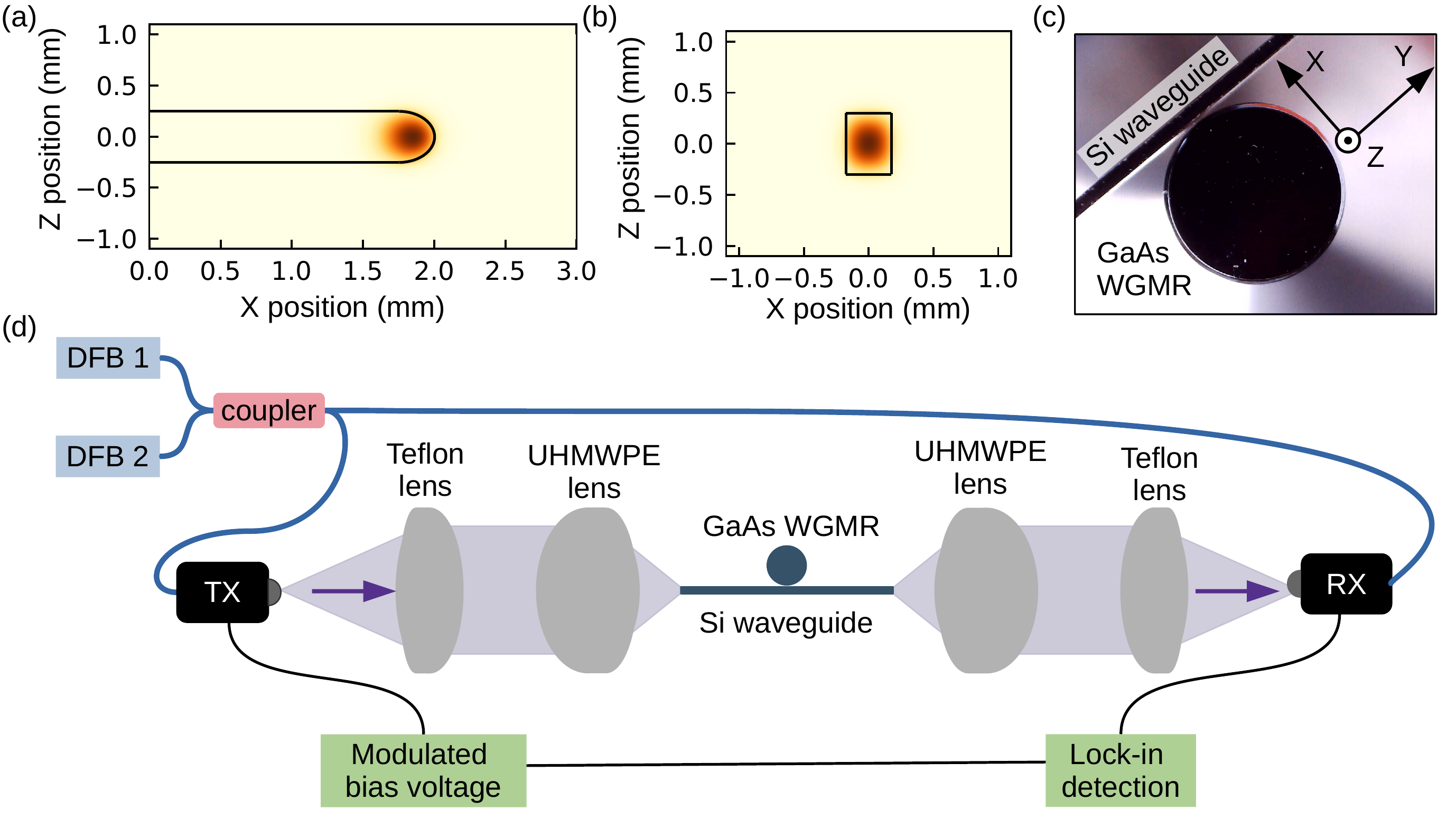}
\caption{Normalised electric field strength at 304.97 GHz in (a) a 2-mm-radius GaAs WGMR disk, and (b) a \qtyproduct{640x350}{\micro m} Si waveguide cross-section. (c) Microscope image of the GaAs WGMR and Si waveguide in the experimental set-up. (d) THz spectrometric system from Toptica Photonics~\cite{deninger_275_2015} with photoconductive antennae (PCA) for the THz transmitter (TX) and detected at the receiver (RX); the radiation is coupled into and from a Si waveguide using UHMWPE lenses~\cite{lo_aspheric_2008} and then coupled between the waveguide and GaAS WGMR by evanescent field coupling. }\label{fig:exptsetup}
\end{figure}

The WGMR to be fabricated was designed with the help of finite element method (FEM) modelling of an axis-symmetric 2D cross-section of the WGMR disk using COMSOL Multiphysics\textsuperscript \textregistered, using the refractive index to be 3.6 and the absorption loss to be below \SI{0.5}{\per\cm}~\cite{yang_study_2014}. The eigenmodes of the GaAs WGMR in different polarisations for the corresponding azimuthal mode number $m$ (i.e., the number of wavelengths that fit in the circumference of the WGMR) are calculated. From the array of vertically-polarised eigenmodes, the free spectral range (FSR) of the resonator is defined by the spectral spacing between consecutive modes of the same mode family. The radius of the resonator was chosen to be \SI{2}{mm} such that the FSR is \SI{6.77}{GHz} at around \SI{300}{GHz}. The cross-section of the normalised electric field strength of the vertically-polarised mode at \SI{304.97}{GHz} and $m=39$ is shown in Fig.\ref{fig:exptsetup}(a).

The GaAs WGMR was fabricated from a \SI[separate-uncertainty = true]{500(20)}{\micro m}-thick wafer with orientation (001) from American Elements\textsuperscript \textregistered. The wafer was characterised by the crystal manufacturer to have a resistivity of up to \SI{4.4e8}{\ohm.cm}. The resistivity goes inversely with the absorption of the material in the THz domain, and therefore high resistivity is necessary to fabricate high quality microresonators~\cite{vogt_terahertz_2022}. A disk was drilled out of the wafer, and was then mounted on a brass rod and spun on a lathe, while a single-point diamond tool was used to cut the edge of the spinning disk. The angle of the diamond tool with respect to the spinning wafer edge (i.e.\ the rake angle) was roughly \ang{40}~\cite{chen_fundamental_2020}. The disk was then polished by hand using \SI{1}{\micro\meter} diamond slurry on a tissue until the scratches on the surface due to the cutting process are smoothed out. The resulting GaAs disk WGMR, \SI[separate-uncertainty = true]{1.997(0.010)}{\mm} in radius and \SI[separate-uncertainty = true]{508(9)}{\micro m} in thickness, was mounted on an aluminium rod with a narrow neck of \SI{0.5}{\mm} in radius to be held in the experimental set-up. 

\subsection{Design and fabrication of the silicon waveguide}
In order to excite the THz modes in the GaAs WGMR, one needs to design and fabricate a waveguide that supports modes that are phase-matched to the WGMR modes. From the azimtuhal mode number $m$ and the eigenfrequency $f$, the effective modal index of the different modes in the WGMR with radius $R$ can be calculated as: $ n_{eff}=m c/(2\pi  R  f)$, where $c$ is the speed of light. Since the effective index of the WGMR modes (for $m=38$, $f=\SI{304.97}{GHz}$, $n_{eff}=2.974$) is close to the refractive index of Si (3.42~\cite{bolivar_measurement_2003}), we identified Si as a good material candidate for the waveguide. 2D FEM modelling of the cross-section of a cuboidal Si waveguide was done using COMSOL Multiphysics\textsuperscript \textregistered, which calculates the effective index in the waveguide. The dimensions of the waveguide were optimised to match the effective indices in both structures, as is required for phase-matching. The refractive index and absorption in Si are taken to be 3.42 and \SI{0.002}{\per\cm} respectively~\cite{bolivar_measurement_2003}. 

The required Si rod waveguide was found to have dimensions of \SI{640}{\micro m} in height and \SI{350}{\micro\meter} in width for exciting vertically-polarised modes in the WGMR. It was fabricated by cutting a Si wafer of thickness \SI{350}{\micro\meter} with the laser micromachining system mentioned in~\cite{vogt_subwavelength_2020}. The system used \SI{20}{W} \SI{120}{fs} pulses centred at \SI{1030}{nm} at a repetition rate of 20 kHz with the stage moving at a speed of \SI{20}{\mm\per\second}. Figure \ref{fig:exptsetup}(b) shows the cross-section of the normalised electric field strength at \SI{304.97}{GHz} in the sub-wavelength waveguide. A 6-cm long piece of fabricated waveguide was mounted on a 5-axis stage to control the height and axial distance on both ends as well as to bring it into the focal planes of the in-coupling and out-coupling lenses. Once the waveguide was aligned in the experimental set-up shown in Fig.~\ref{fig:exptsetup}(d), the WGMR was mounted on a translational stage to bring it closer to the waveguide. 

\subsection{Characterization of the GaAs WGMR}
\begin{figure}[h!]
\centering\includegraphics[width=14cm]{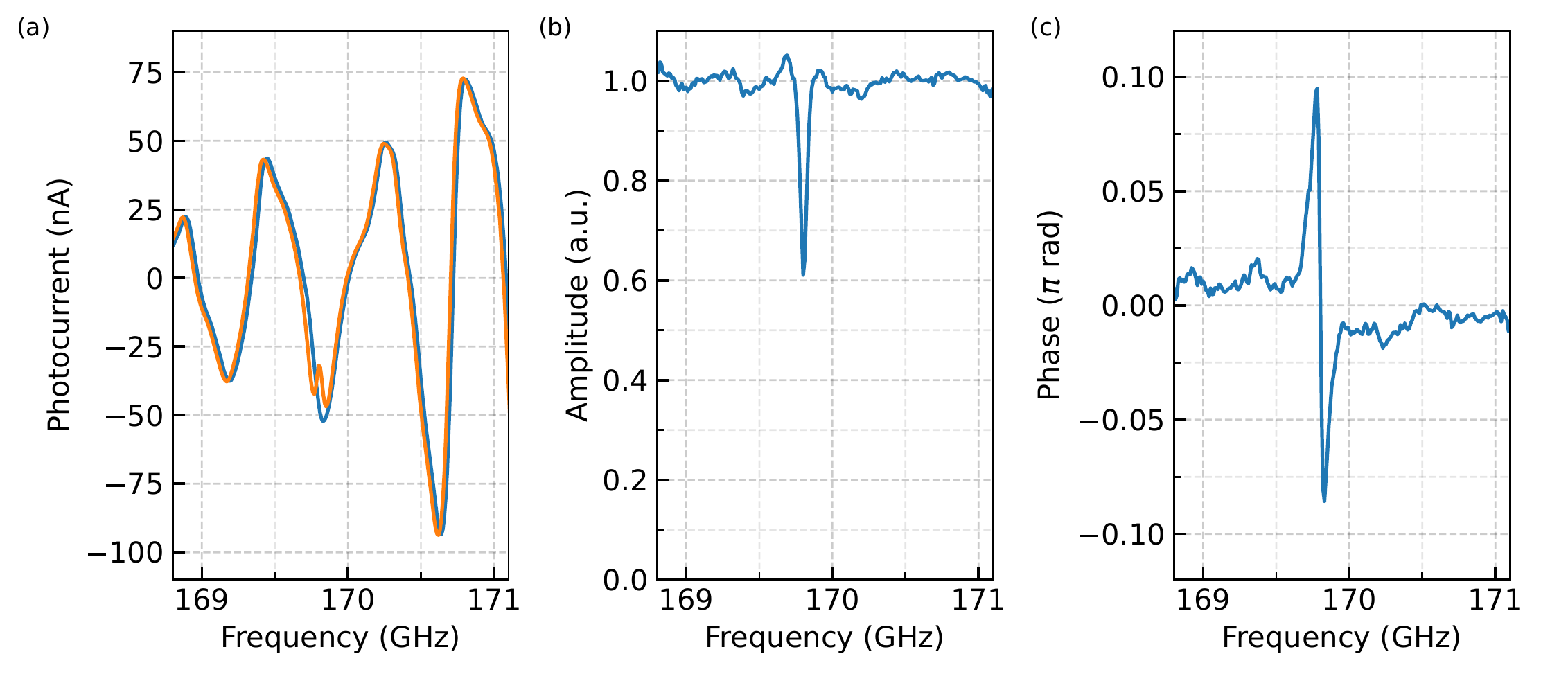}
\caption{(a) Photocurrent signal from \qtyrange{168.5}{171.2}{GHz} with (sample:orange) and without (reference:blue) the WGMR coupled to the waveguide. (b) Amplitude and (c) phase due to the sample from Hilbert transform analysis.}\label{fig:photocurr_amp_phase}
\end{figure}

The set-up for characterizing the Q factor of the resonator modes by coupling the radiation into the WGMR and exciting modes in it is shown in Fig.~\ref{fig:exptsetup}(d). The commercially-available frequency domain THz spectrometric system from Toptica Photonics~\cite{deninger_275_2015} was used. The THz radiation is generated using a pair of lasers at telecom wavelengths whose frequency separation gives the frequency of the THz radiation being emitted from the photoconductive antenna (PCA) on the left shown as TX (transmitter), and detected coherently as a photocurrent at the antenna on the right labelled RX (receiver). The frequency was swept from \qtyrange{150}{380}{GHz}. The Teflon lenses are used to collimate and focus the beam from and to the antennae. A pair of ultra-high-molecular-weight polyethylene (UHMWPE) lenses~\cite{lo_aspheric_2008} are used to couple the radiation to and from a sub-wavelength Si rod waveguide, from which by evanescent field coupling, the radiation in the waveguide excites modes in the WGMR. 

The coherent frequency-domain spectroscopy followed by data analysis using the Hilbert transform~\cite{vogt_high_2017} involves a two-step measurement. First, a reference measurement was taken while the WGMR was far away and not coupled to the waveguide. This smoothly oscillating photocurrent is shown in blue in Fig.~\ref{fig:photocurr_amp_phase}(a). Then, a sample measurement was taken with the WGMR close to and coupled to the waveguide. As can be seen in the orange curve in Fig.~\ref{fig:photocurr_amp_phase}(a), a small kink appears at \SI{169.8}{GHz} due to the coupling of the resonant radiation into the WGMR. A Hilbert transform of the photocurrent signal gives an instantaneous amplitude and phase corresponding to the reference and sample measurements. Taking the ratio of the amplitudes, the amplitude effect of the WGMR can be seen as a dip at the resonant frequency [Fig.~\ref{fig:photocurr_amp_phase}(b)]. The difference in the phases shows a sharp change in phase at the same frequency [Fig.~\ref{fig:photocurr_amp_phase}(c)]. This coherent detection thus provides us additional information with the phase profile: when the system is undercoupled, the shape of the phase shift is as shown in Fig.~\ref{fig:photocurr_amp_phase}(c); and when it is overcoupled, the phase shift would be a jump of 2$\pi$~\cite{bogaerts_silicon_2012}. 

\section{Results and discussion}
The THz spectroscopy system was scanned across different spectral ranges within \qtyrange{150}{380}{GHz}. Reference and sample measurements are acquired as described before, and from here onward, we show only the amplitude and phase obtained from the photocurrent measurements. 

\begin{figure}[h!]
\centering\includegraphics[width=13cm]{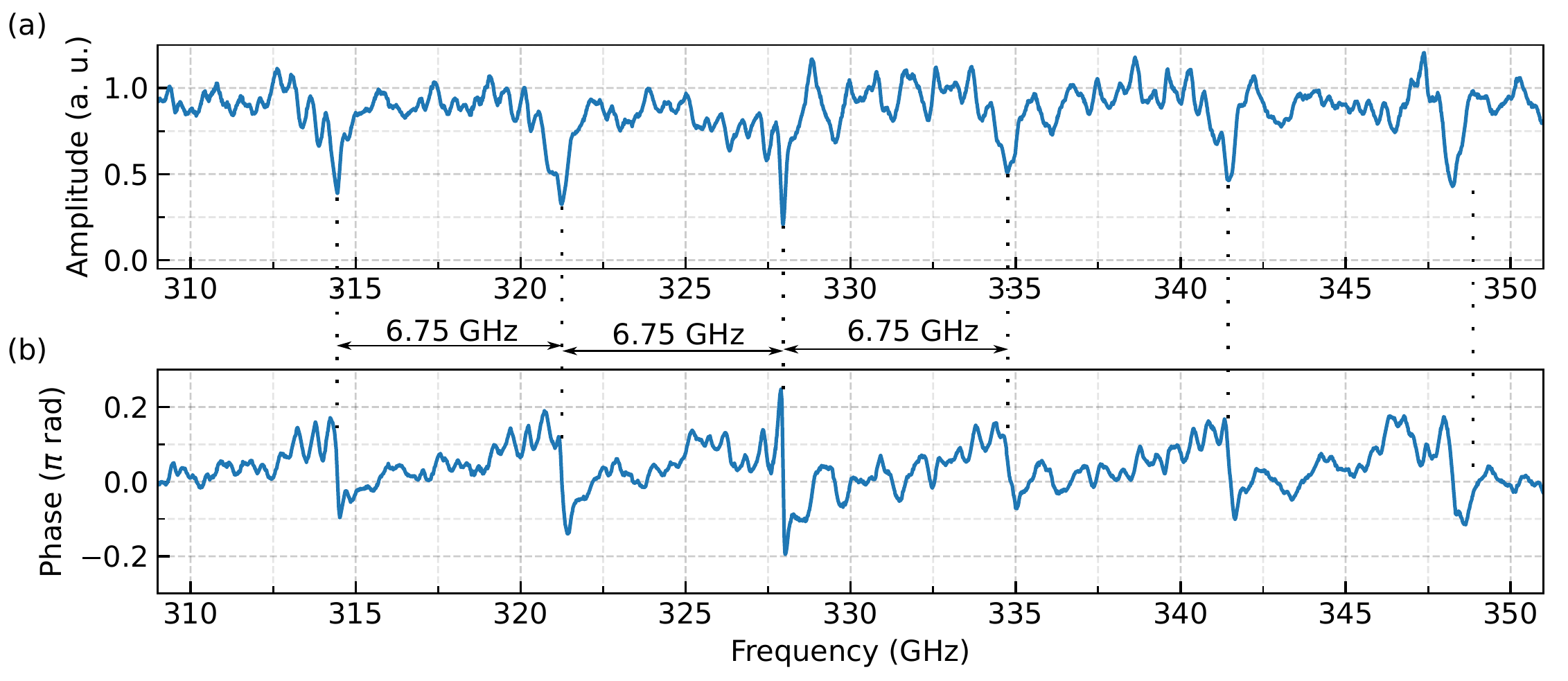}
\caption{(a) Amplitude and (b) phase profiles showing the modes in the GaAs disk resonator in the range from \SI{310}{GHz} to \SI{350}{GHz}.\label{fig:broadscan_298_382}}
\end{figure}

As can be seen in Fig.~\ref{fig:broadscan_298_382}, there are multiple dips in the amplitude spectrum which are spaced from each other by \SI{6.75}{GHz} within the range from \qtyrange{310}{350}{GHz}. This corresponds to the FSR expected from the FEM in this frequency range, if the refractive index of the material in the FEM simulations is modified from 3.6~\cite{yang_study_2014} to 3.65. These modes are identified by comparison to the FEM simulations to be the vertically-polarised fundamental modes, i.e., single-lobed in radial and polar directions.

\begin{figure}[h!]
\centering\includegraphics[width=13cm]{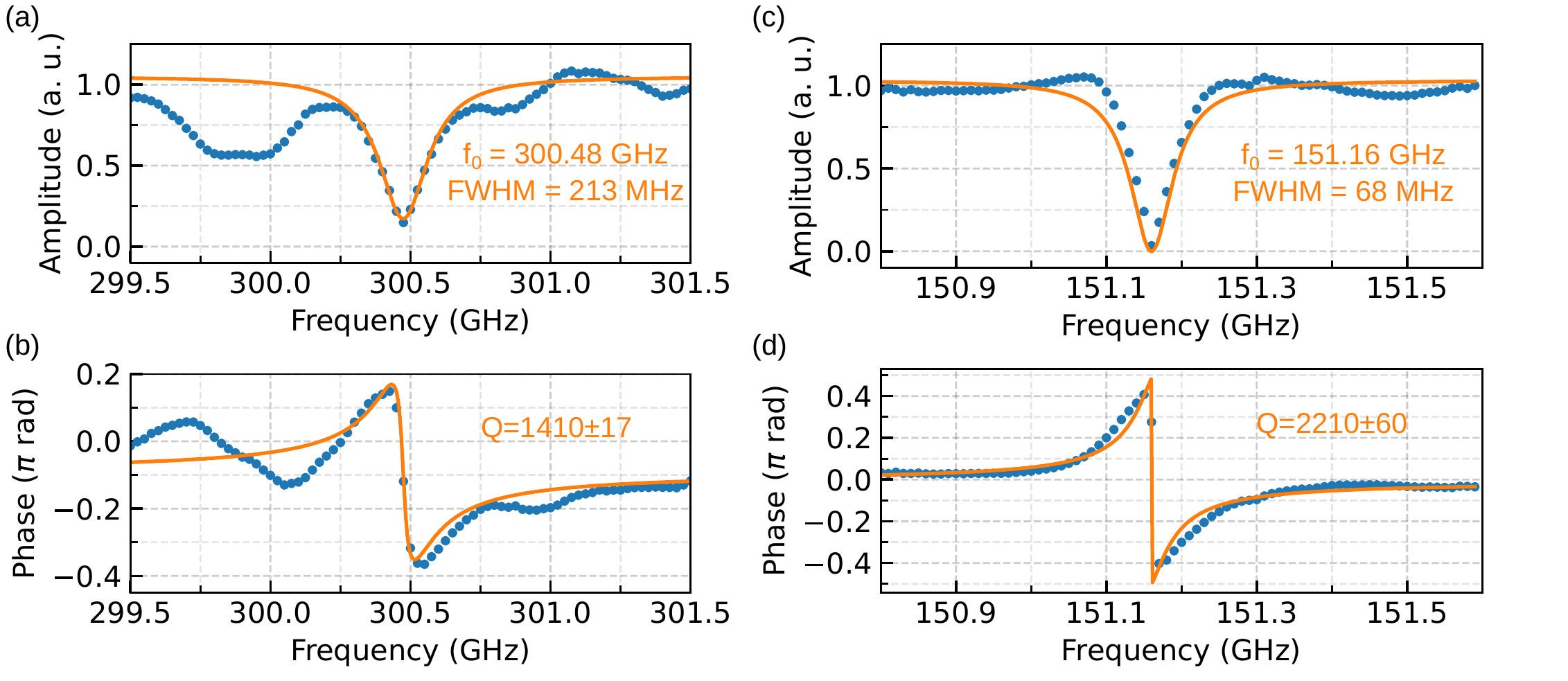}
\caption{Simultaneous fitting of the (a) Amplitude and (b) phase profiles at \SI{300.5}{GHz} and \SI{151.2}{GHz} to obtain to a $Q$ of $\sim1400$ and $\sim2200$ respectively.\label{fig:fitting_f15116_Q2212_Q04494_f300480_Q1414_Q02357}}
\end{figure}

Adjusting the coupling such that a mode at \SI{300.5}{GHz} was undercoupled, a scan with a smaller frequency step size of \SI{5}{MHz} was taken to plot and fit the \SI{300.5}{GHz} mode. A linewidth of \SI{213}{MHz} was obtained by the simultaneous fitting of both the amplitude [Fig.~\ref{fig:fitting_f15116_Q2212_Q04494_f300480_Q1414_Q02357}(a)] and phase [Fig.~\ref{fig:fitting_f15116_Q2212_Q04494_f300480_Q1414_Q02357}(b)] profiles of the mode. This corresponds to a $Q$ of $1.40\text{k}\pm0.02\text{k}$ at \SI{300.5}{GHz}. 

In order to compare the $Q$ at lower frequencies, the system was adjusted to undercouple modes at around \SI{150}{GHz} (which would be overcoupled when measuring undercoupled modes at \SI{300}{GHz}). On simultaneous fitting of the amplitude and phase profiles at \SI{151.2}{GHz}, a linewidth of \SI{68}{MHz} was obtained as shown in Fig.~\ref{fig:fitting_f15116_Q2212_Q04494_f300480_Q1414_Q02357}(c)-(d), which gives a $Q$ of $2.20\text{k}\pm0.06\text{k}$ at \SI{151.2}{GHz}.

As can be seen from the amplitude in both cases, there was significant absorption (the amplitude at resonance is $<25\%$). This indicates that the $Q$ was a loaded $Q$, i.e.\ the waveguide has an effect on the absorption losses in the system. 

\begin{figure}[h!]
\centering\includegraphics[width=13cm]{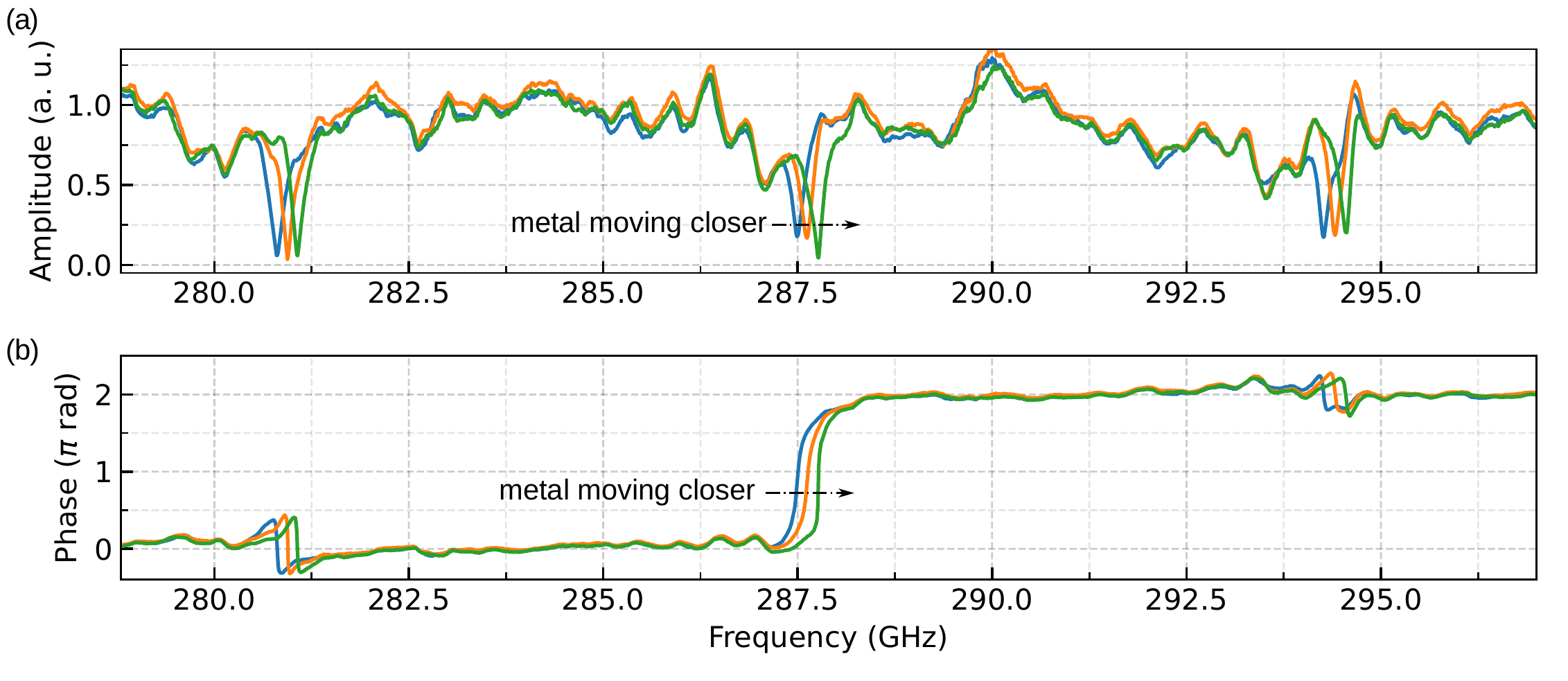}
\caption{(a) Amplitude and (b) phase profiles showing the blue-shifting of the THz modes as an aluminium piece moves closer to the WGMR (blue to orange to green).\label{fig:metal_tuning_280_295}}
\end{figure}

A useful feature of such THz microresonators is that the frequency position of the WGMR modes can be adjusted by introducing a piece of metal into the vicinity of the THz mode~\cite{vogt_anomalous_2019}. Here, we used a block of aluminium and brought it about \SI{0.2}{mm} away from the edge of the GaAs disk. The coupling of the system was adjusted such that we see one overcoupled mode at $\sim\SI{287.5}{GHz}$ between two undercoupled modes at \SI{281}{GHz} and at \SI{294}{GHz}. As we moved the aluminium piece closer by a few micrometers, we saw the THz modes shifting from \qtyrange{280.81}{281.07}{GHz}, \qtyrange{287.49}{287.77}{GHz} and \qtyrange{294.26}{294.56}{GHz} corresponding to the blue to the green curves. The loaded Q factor improves slightly with the blue-shifting, eg. it goes from $975\pm18$ at \SI{280.81}{GHz} to $1032\pm21$ at \SI{281.07}{GHz}, and from $1341\pm24$ at \SI{287.5}{GHz} to $1750\pm78$ at \SI{287.77}{GHz}. This can be attributed to a slight change in coupling strength.

Such tunability of around \SI{0.3}{GHz} can be very useful to relax the stringent requirements during fabrication of the WGMR, since the positions of the THz modes in a WGMR depend strongly on its physical dimensions. This is particularly important for efficient implementation of nonlinear optical processes in such microresonators since the dimensions of the WGMRs play an important role in ensuring phase-matching between THz modes and optical modes as discussed, eg., in~\cite{abdalmalak_integrated_2022,santamaria-botello_sensitivity_2018}. For instance, a \SI{10}{\micro m} change in radius of the disk WGMR results in a shift in the FSR of \SI{0.04}{GHz}, and a \SI{20}{\micro m} change in thickness of the disk WGMR causes a \SI{0.02}{GHz} change in the FSR.

\section{Conclusion}
We presented a disk-shaped GaAs WGMR of radius \SI{2}{mm} which was fabricated by single-point diamond-turning on a lathe followed by polishing with \SI{1}{\micro m} sized diamond slurry. The resonator was characterised using a THz spectroscopy scheme and found to have quality factors of 2200 at $\sim\SI{150}{GHz}$ and 1400 at $\sim\SI{300}{GHz}$. We showed that the frequency positions of the modes can be tuned slightly by introducing a metal in their vicinity. This additional degree of freedom could eliminate the need for iterative fabrication steps to achieve the exact dimensions required for phase-matching of THz modes involved in nonlinear optical processes in the resonator. 

Our evaluation of the GaAs microdisk WGMR shows that it can be a major component in the rapidly developing field of nonlinear THz photonics, especially as part of on-chip platforms for nonlinear generation and detection schemes. For instance, the THz-photonics sources based on optical-to-THz wavelength converters via difference frequency generation~\cite{leon_thz-photonics_2022}, as well as transceivers based on the reverse process can be efficiently realised at low threshold powers in resonance cavities like WGMRs, as has been demonstrated in other resonant designs before~\cite{shijia_cascaded_2020, zeng_efficient_2018}, including WGMRs~\cite{andronico_integrated_2008, sinha_tunable_2016}. 

\begin{backmatter}
\bmsection{Funding} This project is funded by an MBIE Endeavour Fund - Smart Ideas
(UOOX2106) and MBIE Catalyst:Seeding (CSG-UOO2002) in New Zealand.

\bmsection{Disclosures} The authors declare no conflicts of interest.

\bmsection{Data availability} Data underlying the results presented in this paper are available at GaAs WGMR for THz photonics, M. I. Suresh, H. G. L. Schwefel, D. W. Vogt, Zenodo 2023 (DOI: 10.5281/zenodo.7969065).

\bmsection{Acknowledgements} The authors would like to acknowledge the help of Thomas Haase at Photon Factory Auckland for fabricating the silicon waveguides used in this work.

\end{backmatter}

\bibliography{references}

\end{document}